# Alloying behavior of wide band gap alkaline-earth chalcogenides

Samantha L. Millican,[1*] Jacob M. Clary,[1] Christopher J. Bartel,[1] Nicholas R. Singstock,[1] Aaron M. Holder,[1,2,3] and Charles B. Musgrave[1,2,3,4*]

[1]Department of Chemical and Biological Engineering, University of Colorado, Boulder, Colorado 80309, USA

[2]Materials and Chemical Science and Technology Center, National Renewable Energy Laboratory, Golden, CO 80401, USA

[3]Renewable and Sustainable Energy Institute, University of Colorado Boulder, Boulder, CO 80309, USA

[4]Materials Science and Engineering Program, University of Colorado Boulder, Boulder, CO 80309, USA

[*]Correspondence: charles.musgrave@colorado.edu and samantha.millican@colorado.edu

## 1    Abstract

Alloying is a powerful tool for tuning materials that facilitates the targeted design of desirable properties for a variety of applications. In this work, we provide a comprehensive investigation of the synthetic accessibility and electronic properties of nine alkaline-earth chalcogenide anion alloys ($CaS_{1-x}O_x$, $CaS_{1-x}Se_x$, $CaS_{1-x}Te_x$, $SrS_{1-x}O_x$, $SrS_{1-x}Se_x$, $SrS_{1-x}Te_x$, $MgS_{1-x}O_x$, $MgS_{1-x}Se_x$, and $MgS_{1-x}Te_x$). We show that isostructural alloying within the rock salt structure is favored for all systems except $MgS_{1-x}Te_x$, which is predicted to be a heterostructural alloy between the rock salt and wurtzite structures. Alloys of S and Se are shown to be readily accessible for all cations with low miscibility critical temperatures, enabling continuous tuning of electronic properties across this composition space. Alloys of S and Te have higher critical temperatures but may be accessible through non-equilibrium synthesis strategies and are predicted here to have desirable electronic properties for optoelectronics with wide band gaps and lower effective masses than alloys of S and Se. Anion alloying in $MgS_{1-x}Te_x$ stabilizes the wurtzite structure across a significant fraction of composition space, which may make it of particular interest as a transparent conducting material due to its lower effective masses and a higher band gap than the rock salt structure. Zero-point corrected random phase approximation (RPA) energies were computed to resolve the small polymorph energy differences of the Mg compounds and are shown to be critical for accurately describing the thermodynamic properties of the corresponding alloys.

## 2    Introduction

Wide band gap (WBG) semiconductors are a broad class of materials which are widely used in commercially produced electronic devices as well as in emerging energy applications, such as photovoltaic (PV) solar cells and photoelectrochemical (PEC) water splitting devices.[1-7] These materials enable the use of smaller, more efficient power electronic components than their silicon-based counterparts due to their higher band gaps, melting points, and thermal conductivities.[6,7] The most well-studied classes of WBG semiconductors are nitride (e.g. GaN), carbide (e.g. SiC), and group VI chalcogenide ($Ch$ = O, S, Se, Te) materials.[5] Although group VI chalcogenide compounds have been synthesized with a wide variety of cations, the simplest and most common WBG chalcogenide semiconductors are formed via the incorporation of group II (e.g. Mg, Ca, or Sr) and $II_B$ (e.g. Zn or Cd) cations.[5] The wide band gap and low dielectric constants of binary alkaline-earth chalcogenides (AECs; AE$Ch$, where AE = Mg, Ca, Sr) in particular have led to their application in blue- and ultraviolet-wavelength optoelectronics,[8-14] luminescent devices,[15-19] microelectronics,[20-22] and magneto-optical devices.[23] Because many binary AECs exhibit desirable optoelectronic and structural properties and similar crystal structures, alloying between their compositions presents a compelling strategy to further improve their properties.[24,25]

Cation and anion alloying have been used in a variety of materials systems, including II-VI chalcogenide semiconductors, with enormous success to design materials with targeted properties tuned for specific technological applications.[26-32] However, despite their many desirable properties, the few studies examining the benefits of alloying AECs have primarily focused on isostructural cation alloys between two compounds with the rock salt structure.[12,24,25,33-35] Theoretical studies of the isostructural rock salt cation alloys of $Mg_xSr_{1-x}Ch$ and $Mg_xCa_{1-x}Ch$ ($Ch$ = S, Se, Te) showed that the band gap could modulated by up to 0.9 eV[24,25] Similarly, theoretical[12] and experimental[33] studies of the structural, electronic, and photoluminescence properties of the isostructural rock salt $Ca_xSr_{1-x}S$ alloy demonstrated that its band gap depends nonlinearly on composition and attributed this effect primarily to volume deformation effects. A theoretical study investigated anion alloying in the $MgS_xSe_{1-x}$, $MgS_xTe_{1-x}$, and $MgSe_xTe_{1-x}$ systems, and although only the rock salt structure was considered and a comprehensive assessment of the alloy stability was not conducted, this study predicted that the band gap could be modulated by up to 2.8 eV using anion alloying.[35]

Although the rock salt structure has been the focus of previous AEC alloy studies, binary AECs have been extensively characterized in a number of polymorphs, including the rock salt, zinc blende, wurtzite, and nickel arsenide structures.[8,15,23,36-43] Ca and Sr chalcogenides have been shown experimentally and computationally to adopt the rock salt crystal structure, but the ground-state structures of the magnesium chalcogenides have not been conclusively determined because the polymorphs exhibit extraordinarily small energy differences below the resolution of all but the most accurate computational methods.[22,36,44-49] For instance, a number of computational studies have predicted different ground-state structures for MgTe and MgSe depending on the density functional theory (DFT) exchange-correlation functional used.[8,15,37,50-55] Specifically, the local density approximation (LDA) functional[56] predicts that MgSe (MgTe) has a rock salt (nickel arsenide) ground-state crystal structure, while functionals using the generalized gradient approximation (GGA)[57] predict a wurtzite (wurtzite) ground-state.[8,15,22,39,51] Although GGA functionals are generally more accurate than the LDA functional for predicting the relative stabilities of bulk phases,[58-61] both approximations predict that the MgSe and MgTe polymorph energy differences are typically within 10-40 meV/atom and as low as 1 meV/atom, well below the resolutions of LDA or GGA. The nearly degenerate stabilities of the MgSe and MgTe crystal structures suggest that heterostructural alloys are possible and that a more accurate functional is required to resolve their ground-state structures and accurately predict the resulting alloy properties.

The objective of the present study is to determine the structural, electronic, and thermodynamic properties of nine AE(S,$Ch$) anion alloys, where AE=Mg, Ca, Sr and $Ch$ = O, Se, Te, in the rock salt (RS), zinc blende (ZB), wurtzite (WZ), nickel arsenide (NA), and hexagonal boron nitride (BN) crystal structures at ambient pressure. The meta-GGA SCAN (strongly constrained and appropriately normed)[62] functional and random phase approximation (RPA)[63] were utilized to probe the impact of functional choice and more accurately predict the energies of Mg$Ch$ polymorphs and their alloys. The predicted trends in the electronic properties of the AEC alloys demonstrate the ability to continuously tune the band gaps and effective masses of these materials through anion alloying. We show that this approach provides a fundamental understanding of the polymorph stability in these compounds and hypothesize that the trends in thermodynamic and electronic properties we predict can guide future experimental studies towards the rational design of new alloys.

### 3    Results and Discussion

#### 3.1  Calcium- and Strontium-Containing Alloys

Alloy mixing enthalpies for the Ca compounds—CaS$_{1-x}$O$_x$, CaS$_{1-x}$Se$_x$, and CaS$_{1-x}$Te$_x$—are shown in Figures 1a, c, and e. The analogous results for Sr compounds—SrS$_{1-x}$O$_x$, SrS$_{1-x}$Se$_x$, and SrS$_{1-x}$Te$_x$—are shown in Figure S1 of the Supplemental Information (SI) and display similar trends to those of the Ca compounds. The RS crystal structure is the predicted ground-state across all composition ranges for each of the Ca and Sr alloys, consistent with experimental observations.[33,64-68] The NA structure is the second lowest energy structure for the sulfide, selenide, and telluride compounds and BN is the second lowest energy structure for the oxides. The lowest energy structures for these compounds are consistent with established crystal structure trends in binary compounds based on the ratio of their cation to anion ionic radii.[69] Specifically, for cation-anion radii ratios between 0.41 and 0.73, octahedral coordination is expected to be present, thus producing the RS and NA structures, while for ratios between 0.23 and 0.41, tetrahedral coordination is expected, thus producing WZ and ZB structures. The cation-anion radii ratios in Ca and Sr range from 0.45 – 0.71 and 0.51 – 0.81, respectively, as the anion decreases in size from Te to O. Only SrO, with a cation-anion radii ratio of 0.81, falls slightly outside of the ideal octahedral range, preferring larger coordination numbers based on this radii ratio rule.

The $CaS_{1-x}Se_x$ and $SrS_{1-x}Se_x$ RS alloys have $\Omega$ = 77 meV and 64 meV, respectively, which are the lowest computed $\Omega$ values of any RS Ca and Sr alloys. The low $\Omega$ values for the ground-state RS structure indicate that the (S,Se) alloys will readily form in these systems, enabling tuning of the material properties through anion alloying across the composition space. Furthermore, the $\Omega$' values for both of these compounds are very small and result in nearly symmetric enthalpy of mixing curves.

The $CaS_{1-x}O_x$ and $SrS_{1-x}O_x$ RS alloys have the largest alloy interaction parameters of the Ca and Sr RS structures, with $\Omega$ = 1365 meV and 1123 meV, respectively, and considerable asymmetry in their enthalpy of mixing curves towards their oxide end-members, as indicated by $\Omega$' = -622 meV and -465 meV, respectively. The large $\Omega$ values indicate the existence of a large driving force for phase segregation (spinodal decomposition) into the binary end-members—CaS/CaO and SrS/SrO. The $CaS_{1-x}Te_x$ and $SrS_{1-x}Te_x$ RS alloys both have intermediate $\Omega$ values of 547 meV and 467 meV, respectively, and moderately large asymmetry in $\Delta H_m(x)$ towards their sulfide end-members with $\Omega$' = 206 meV and $\Omega$' = 164 meV, respectively. We note that for $x$ = 1, the CaTe and SrTe NA structures lie 80 meV/fu and 126 meV/fu higher in energy than their respective RS structures and are the smallest polymorph energy differences computed for any Ca-based or Sr-based compound studied in this work.

These polymorph energy differences are larger than the approximate resolution of the SCAN functional,[70] and the predicted ground-state crystal structures are consistent with experimental observations.[33,64-68]

The temperature-composition (*T-x*) phase diagrams for each of the Ca-containing compounds are shown in Figures 1b, d, and f and for each of the Sr-containing compounds in Figures S1b, d, and f. For the Ca- and Sr-containing compounds, a single RS alloy is the most stable across the entire composition space. This results in a conventional isostructural phase diagram in which the spinodal and binodal curves meet at the miscibility gap critical temperature, leading to relatively narrow metastable regions. Above the miscibility gap critical temperature, a fully thermodynamically miscible alloy is predicted across the entire composition space. From these *T-x* phase diagrams, the effects of the calculated mixing enthalpy on stability and synthesizability become evident. For instance, because *Ω* is small for both the $CaS_{1-x}Se_x$ and $SrS_{1-x}Se_x$ alloys, the miscibility gap critical temperatures are thus also low, 180 °C for $CaS_{1-x}Se_x$ and 105 °C for $SrS_{1-x}Se_x$, indicating that phase pure RS (S,Se) alloys are synthesizable across the entire range of alloy compositions at modest synthesis temperatures. In contrast, because the RS $CaS_{1-x}O_x$ alloy has the highest ground-state alloy interaction parameter ($\Omega$ = 1365 meV) of the Ca and Sr compounds examined in this work, we predicted that it has an extremely narrow stability region up to the melting temperature of the end-members and that highly non-equilibrium growth would be required to realize a solid solution. Indeed, except for the Ca and Sr (S,Se) alloys, all other Ca- and Sr-containing alloys studied here exhibit high $\Omega$ values, suggesting that highly non-equilibrium growth is required for their synthesis. For instance, reactive co-sputtering has produced phase pure alloys across approximately 75% of the $Al_{1-x}Sc_xN$ composition space (x < 0.4 and x > 0.65), despite this system's $\Omega$ being at least 970 meV.[71] Thus, because we predict substantially lower $\Omega$ values for the (S,Te) alloys we expect non-equilibrium growth to be an effective strategy for synthesizing these alloys worthy of future study. We discuss the large difference in alloy interaction parameters between these alloys in more detail below.

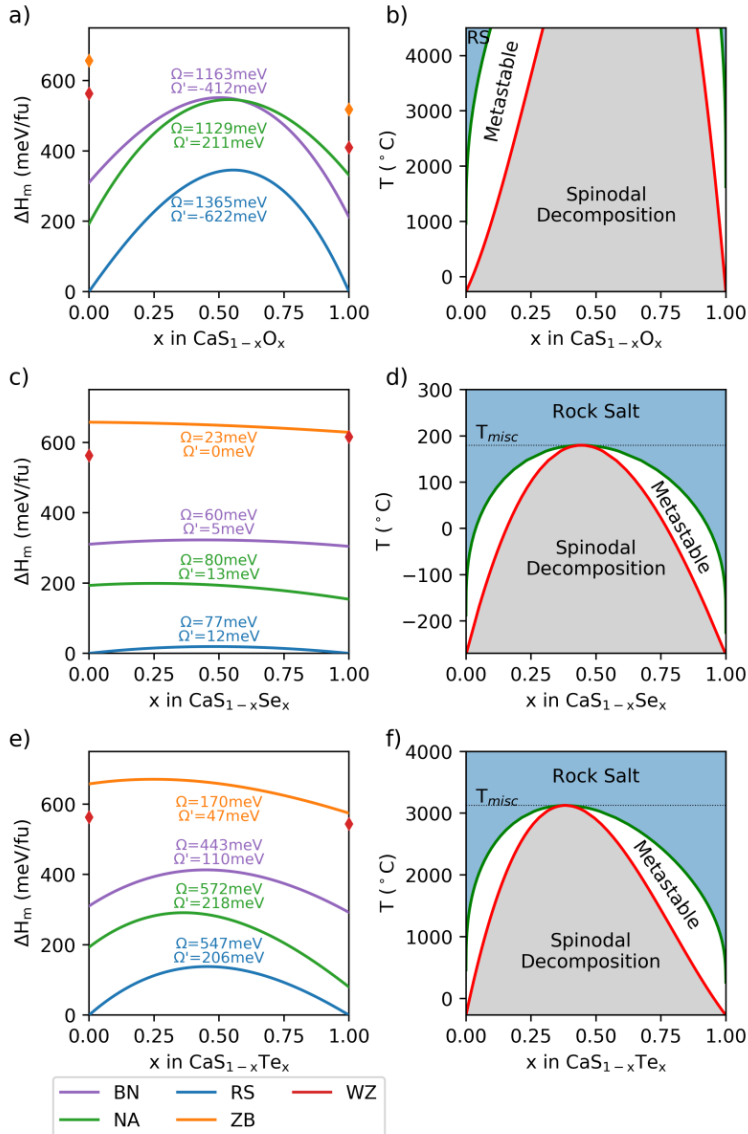

Figure 1. *Thermodynamic stability of Ca compounds*. Calculated mixing enthalpy curves for each of the five polymorphs considered (left) and the resulting temperature-composition phase diagram (right) for calcium-containing alloys a-b) $CaS_{1-x}O_x$, c-d) $CaS_{1-x}Se_x$, and e-f) $CaS_{1-x}Te_x$. The alloy interaction parameters *Ω* and *Ω'* are given for each polymorph in the mixing enthalpy figures. For compounds found to be unstable, only the end-member energies are displayed. In the phase diagrams the spinodal line (red) and binodal line (green) separate the regions of stability (blue), metastability (white), and instability (grey).

The changes in the *GW*-corrected DOS effective masses and *GW*-corrected band gaps over the composition space for the Ca and Sr (S,Se) and (S,Te) RS alloys are shown in Figure 2. The oxide structures are omitted due to their low miscibility. Our results show that for these compounds, anion alloying in Ca and Sr chalcogenides can be used to modulate the resulting compound's band gap nearly continuously from 4.1 to 5.4 eV and from 3.8 to 4.7 eV, respectively. For both the Ca and

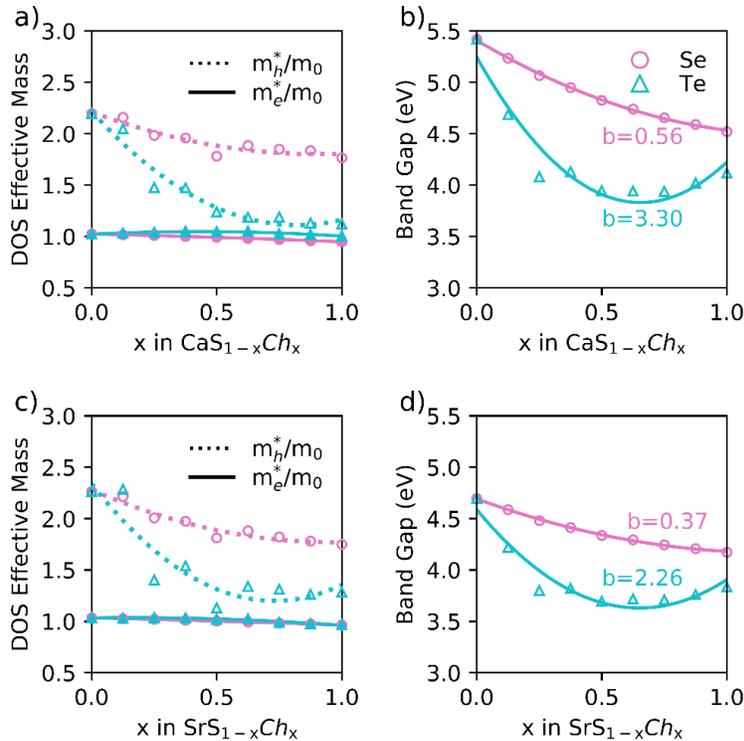

Figure 2. *Electronic properties of Ca and Sr compounds.* Trends in the calculated *GW* DOS effective mass of holes and electrons and effective *GW* band gaps over composition space for Ca compounds (a-b) and Sr compounds (c-d) for the ground-state RS structure. Selenides are shown as pink circles and tellurides are shown as teal triangles. Oxides are omitted for simplicity due to their low miscibility. For each compound the band gap bowing parameter (*b*, in eV) is given.

Sr chalcogenides, the (S,Se) alloys have larger band gaps and smaller optical bowing coefficients than the (S,Te) alloys, but the band gaps of all these systems are > 3.1 eV, making them transparent. These band gaps are larger than those of many other binary chalcogenide semiconductors such as Zn*Ch*, Mn*Ch*, and Cd*Ch*,[5] making them potentially useful transparent conductive materials. The effective masses of a material can be used to estimate carrier mobility, where low electron effective mass, $m_e^*$, is required for for n-type conductivity and low hole effective mass, $m_h^*$, is required for p-type conductivity.[5] Anion alloying does not significantly affect $m_e^*$, which is fairly constant at ~1 for the Ca and Sr compounds. In contrast, $m_h^*$ changes significantly depending on composition and can be continuously tuned from 1.1 to 2.3 through anion alloying. CaTe and SrTe exhibit the lowest $m_h^*$ at 1.1 and 1.3, respectively, indicating their potential use as high-mobility p-type transparent conducting materials.

From this study of Ca and Sr-based AECs, (S,Se) and (S,Te) alloys have been identified as promising materials for optoelectronic applications. The low miscibility gap temperatures of the (S,Se) alloys enable the synthesis of thermodynamically stable, single-phase alloys at relatively low temperatures, and provide both band gap and $m_h^*$ tunability. Although the (S Te) alloys have higher miscibility gap temperatures, their low $m_h^*$ may make them more promising as p-type transparent conducting materials.

### 3.2 Magnesium-Containing Alloys

As discussed above, Mg chalcogenides have very small predicted polymorph energy differences and, consequently, have been observed experimentally in various structures. MgS and MgSe have been synthesized in the WZ, ZB, and RS structures while MgTe has been synthesized in the WZ, ZB, RS, and NA structures.[46,47,72-77] Because correct energy ordering of the end-member crystal structures is critical for predicting alloy properties, we compared the performance of the

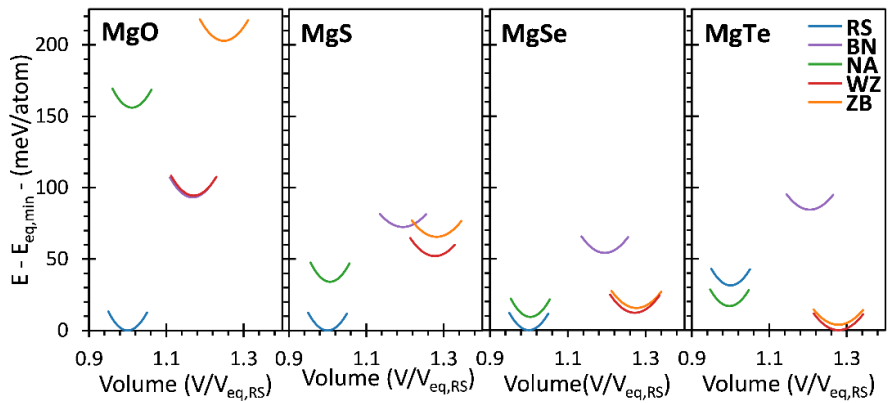

Figure 3. *Polymorph stability of Mg compounds.* Relative polymorph energies calculated using RPA+ZPE as a function of the fractional volume (V/V$_{eq,RS}$) for MgO, MgS, MgSe, and MgTe in the BN, NA, RS, WZ, and ZB.

PBE and SCAN DFT functionals to the polymorph ordering predicted using the RPA with ZPE. When coupled with exact exchange, the RPA has been shown to predict lattice constants, polymorph energy differences, and formation enthalpies that are in better agreement with experiment than the LDA and PBE functionals.[78-81] The ZPE-corrected RPA energy-volume curves are shown in Figure 3. These calculations predict that MgO, MgS, and MgSe have a RS ground-state structure and MgTe has a WZ ground-state structure. A comparison of the polymorph energy differences and structural parameters computed with PBE, SCAN, and RPA are reported in Tables 1 and S1. For MgO and MgS, PBE, SCAN, and RPA all predict a RS ground-state structure; however, the rank-ordering of the other structures predicted by both PBE and SCAN deviates from that of RPA. For MgS, both PBE and SCAN predict WZ to be the second lowest energy structure at 8 and 23 meV/atom above the ground state (RS), respectively, while RPA predicts NA to be lower in energy than WZ by 20 meV/atom. Furthermore, for MgSe, both PBE and SCAN predict WZ to be the ground-state structure, followed by ZB and RS; however, RPA predicts RS to be the ground-state structure followed by NA and WZ. In MgSe, all structures except for BN are within a range of 25 meV/atom, which is generally too small of an energy difference to be resolved using PBE or SCAN. This difference in the ground-state structure will result in qualitatively different results in the assessment of the thermodynamic stability of the alloys and highlights the importance of using accurate computational methods like RPA for compounds with small energy differences. In MgTe, RPA, PBE, and SCAN all predict WZ to be the ground-state structure followed closely by ZB at 4, 1, and 1 meV/atom, respectively, although these energy differences are all well within the resolution of any DFT functional and also RPA with ZPE. This result contradicts multiple published reports using the LDA functional in which NA was reported as the ground-state structure.[8,15,39] The comparison of the stabilities predicted by these computational methods illustrates the importance of utilizing the highest accuracy methods that are computationally feasible, such as RPA, for calculating polymorph energetics when the energy differences are less than ~40 meV/atom.[70,82]

Table 1. Polymorph energy differences (meV/atom) for MgO, MgS, MgSe, and MgTe in the hexagonal boron nitride (BN), nickel arsenide (NA), rock salt (RS), wurtzite (WZ), and zinc blende (ZB) structures.

|     | MgO | | | MgS | | | MgSe | | | MgTe | | |
| --- | --- | --- | --- | --- | --- | --- | --- | --- | --- | --- | --- | --- |
|     | PBE | SCAN | RPA | PBE | SCAN | RPA | PBE | SCAN | RPA | PBE | SCAN | RPA |
| **RS** | 0.0 | 0.0 | 0.0 | 0.0 | 0.0 | 0.0 | 18.4 | 12.6 | 0.0 | 64.2 | 69.8 | 21.1 |
| **BN** | 37.6 | 62.0 | 98.8 | 32.1 | 48.2 | 75.3 | 42.7 | 48.9 | 57.4 | 80.7 | 93.6 | 76.5 |
| **NA** | 148.8 | 159.0 | 154.2 | 26.5 | 30.3 | 37.0 | 25.2 | 22.7 | 12.4 | 47.1 | 54.1 | 10.1 |
| **WZ** | 37.6 | 115.1 | 99.5 | 7.7 | 22.7 | 59.6 | 0.0 | 0.0 | 20.9 | 0.0 | 0.0 | 0.0 |
| **ZB** | 105.9 | 152.1 | 208.6 | 16.5 | 31.5 | 73.0 | 5.2 | 5.2 | 24.3 | 1.1 | 0.8 | 3.8 |

Thermochemical calculations of the alloy enthalpy of mixing for the Mg alloys, $MgS_{1-x}O_x$, $MgS_{1-x}Se_x$, and $MgS_{1-x}Te_x$, are shown in Figures 4a, c, and e. For the $MgS_{1-x}O_x$ alloy, both the MgO and MgS end-members exhibit a ground-state RS crystal structure. However, the dense packing of the RS structure makes substitution reactions involving differently sized anions unfavorable and results in a very large $\Omega$. This large $\Omega$ in conjunction with the small polymorph energy differences in the end-member compounds enables a structural phase transition between RS and the less-densely packed WZ structure which has lower alloy interaction parameters. However, the crossovers of the RS and WZ enthalpy of mixing curves occur for $\Delta H_m \geq 400$ meV/fu, suggesting that the WZ structure will likely not be realized experimentally. The difference in $\Omega$ between polymorphs of different density has been observed in similar systems, including Mn(Se,Te).[32] Similar to the Ca and Sr (S,Se) alloys, we predict that the $MgS_{1-x}Se_x$ alloy has the smallest magnitude alloy interaction parameters of the three Mg-containing alloys, with $\Omega = 103$ meV and $\Omega' = 16$ meV for the ground-state RS structure, suggesting the potential for continuous compositional tuning in this space. Despite RPA polymorph energy differences as low as 22 meV/fu for $MgS_{1-x}Se_x$, a small $\Omega$ results in the RS structure being the energetic ground state across the full composition space with no predicted phase transitions at ambient pressure. In the $MgS_{1-x}Te_x$ system, MgS has a RS ground-state structure while MgTe has a WZ ground-state structure, resulting in a unique heterostructural alloy with a phase transition at $x = 0.30$. In this compound, the small polymorph energy differences in MgS and MgTe enable a large compositional range ($x > 0.30$) of low-energy WZ (and potentially ZB) structures to be achieved in the AEC family of compounds that is largely dominated by RS structures. This large stability window of the WZ structure is unique to the anion alloying studied here. If cation alloying was attempted within the AEC family, such as in $Mg_xCa_{1-x}Te$ or $Mg_xSr_{1-x}Te$, the high polymorph energies of WZ CaTe and SrTe relative to the RS structure would likely result in an extremely narrow composition range in which WZ would be the ground-state structure in those heterostructural alloys.

The resulting *T-x* phase diagrams for Mg-containing compounds are shown in Figures 4b, d, and f. The large alloy interaction parameters of the $MgS_{1-x}O_x$ alloy result in the narrowest stability regions within the temperature range shown for any of the nine alloys examined. The $MgS_{1-x}Se_x$ phase diagram is also similar to those of the $CaS_{1-x}Se_x$ and $SrS_{1-x}Se_x$ phase diagrams with a moderate miscibility gap critical temperature (< 400 °C). The $MgS_{1-x}Se_x$ alloy system exhibits the largest alloy interaction parameters of the (S,Se) alloys, but a fully thermodynamically miscible single-phase alloy can still be achieved at temperatures above 330 °C across the entire composition space. Due to the heterostructural nature of the $MgS_{1-x}Te_x$ alloy, this system has a unique phase diagram relative to those of the isostructural alloys presented thus far and therefore offers certain advantages over the isostructural alloys. In the case of $MgS_{1-x}Te_x$, alloying occurs between the incommensurate RS and WZ lattices which requires a reconstructive phase transition at $x = 0.30$. This reconstructive phase transition results in a discontinuity in the spinodal curve and a decoupling of the binodal and spinodal miscibility gaps, which creates a large region of metastability in the phase diagram.[29] This opens a novel and intriguing space for future synthesis studies because the stability against composition fluctuations in the heterostructural RS/WZ alloy could enable the synthesis of a single-phase WZ AEC across a wide composition range with tunable properties for optoelectronics and other applications.

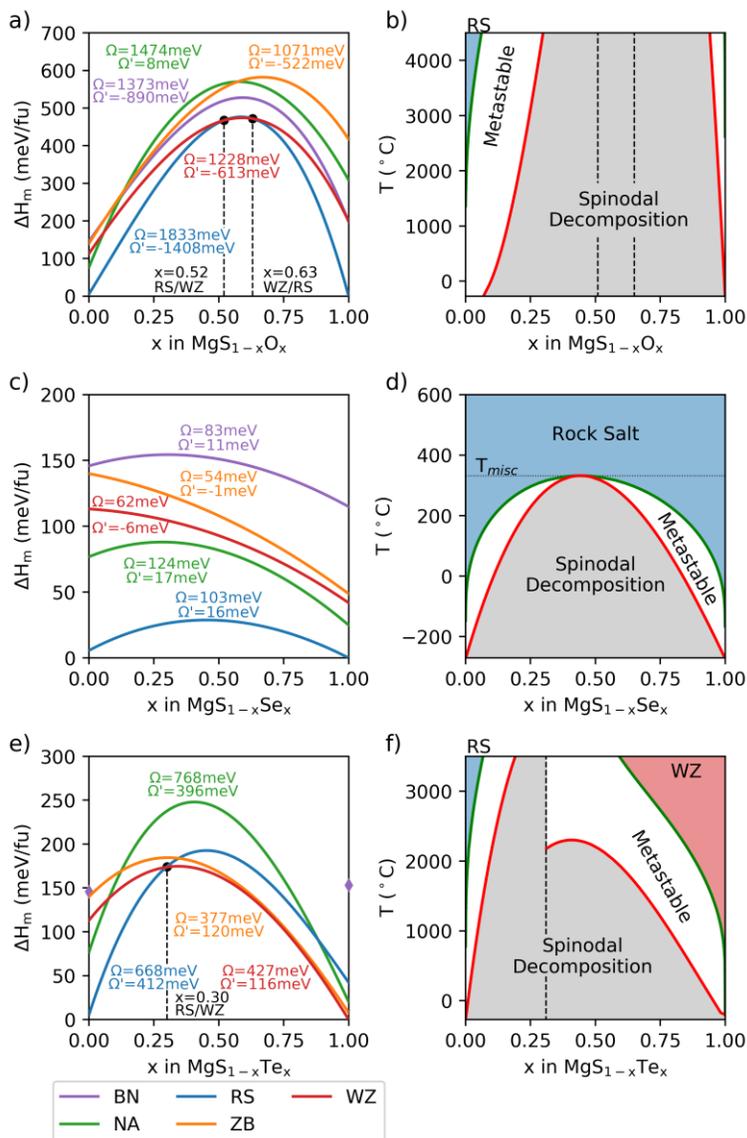

Figure 4. *Thermodynamic stability of Mg compounds.* RPA-corrected enthalpy of mixing curves for each of the five polymorphs considered (left) and resulting temperature-composition phase diagrams (right) for strontium-containing alloys a-b) $MgS_{1-x}O_x$, c-d) $MgS_{1-x}Se_x$, and e-f) $MgS_{1-x}Te_x$. In the enthalpy of mixing figures alloy interaction parameters *Ω* and *Ω'* are given for each polymorph. For compounds found to be unstable, only the end-member energies are displayed. In the phase diagrams the spinodal curve (red) and binodal curve (green) separate the regions of stability (blue), metastability (white), and instability (grey).

Computed *GW*-corrected effective masses and *GW*-corrected band gaps for Mg alloys across the composition space are presented in Figure 5. The oxide alloys are not shown due to their low miscibility. Computed values for the MgS$_{1-x}$Te$_x$ RS, WZ, and ZB structures are all shown. Within the Mg-chalcogenide space, the band gaps can be continuously modulated from 1.1 eV for RS MgTe to 5.7 eV for WZ and ZB MgS, offering a large degree of tunability for targeted applications. The band gaps in the RS Mg(S,Se) and Mg(S,Te) alloys are lower than those in the analogous RS Ca and Sr compounds, and the band gap of RS MgS$_{1-x}$Te$_x$ for $x > 0.15$ is below the approximate 3.1 eV cutoff for transparent materials. However, the band gap for the WZ MgS$_{1-x}$Te$_x$ structures are above 3.1 eV across the composition space, indicating that the WZ structure, which is the ground-state structure for $x > 0.30$, could potentially be used as a transparent conducting material. In contrast to the calculated $m_e^*$ of the Ca and Sr compounds, which were all relatively constant at ~1, $m_e^*$ changes considerably across the composition space in the Mg compounds, as shown in Figure 5b. In particular, WZ and ZB MgS$_{1-x}$Te$_x$ have low $m_e^*$ of ~0.5 at the phase transition point ($x = 0.30$). The WZ structure of MgS$_{1-x}$Te$_x$ also has the lowest $m_h^*$ of any of the Mg compounds studied, as shown in Figure 5a. By utilizing anion alloying in this system to stabilize the WZ structure across a significant fraction of the composition space, low $m_e^*$, low $m_h^*$, and wide band gaps can be achieved which may make them of particular interest as transparent conducting materials for a wide range of applications such as solar cells or electronic devices. Although the miscibility gap temperature in this compound is exceptionally high, the heterostructural nature of the alloy results in a wide metastable region and accessing the WZ structure, even at relatively low S concentrations, may be possible through a highly non-equilibrium growth method.

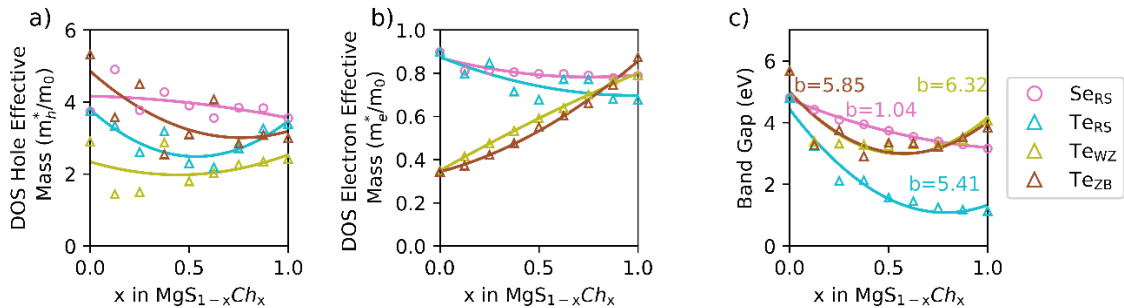

Figure 5. *Electronic properties of Mg compounds.* Trends in the calculated *GW*-corrected DOS effective masses of holes (a) and electrons (b) and *GW*-corrected band gaps (c) over the composition space of RS MgS$_{1-x}$Se$_x$ (pink circles), and RS, WZ, and ZB MgS$_{1-x}$Te$_x$ (teal, olive, and brown triangles, respectively). MgS$_{1-x}$O$_x$ is omitted due to its low miscibility. The band gap bowing coefficient *b* is given for each compound in eV.

### 3.3 Discussion of Periodic Trends

Examining the results for the nine alloy systems studied in this work reveals trends in $\Omega$, miscibility temperature, and band gap with changing cation and anion. Both $\Omega$ and the miscibility temperature decrease across the cationic series Mg > Ca > Sr, correlating with increasing volume, and decrease across the anionic series (S,O) > (S,Te) > (S,Se), correlating with decreasing difference in anionic radii. The computed band gaps increase with increasing anion radii from O to Te, correlating with decreasing electronegativity. The origins of these trends are discussed in further detail in the SI, but these trends can be utilized to extend our understanding of additional systems. Although this study was conducted utilizing S as the common anion between all alloys, we propose that the trends discussed above can be used to predict the relative properties of other anion-anion alloy pairs. For example, the differences in the ionic radii for (S,O), (S,Te), (Se,O) and (Se,Te) are 0.44, 0.37, 0.58 and 0.23 Å, respectively. As a result, we predict that (Se,O) alloys will exhibit higher alloy interaction parameters and miscibility temperatures than (S,O) alloys with the same cation, while (Se,Te) will exhibit lower alloy interaction parameters and miscibility temperatures than (S,Te) alloys with the same cation. Because (S,Te) alloys have been identified in this work as interesting WBG semiconductors due to their wide band gaps and low effective masses, but suffer from higher miscibility temperatures than (S,Se) alloys, (Se,Te) alloys might serve as a more easily synthesized alternative alloy that maintains the desirable and tunable electronic properties of (S,Te) alloys.

As was noted previously, Mg-containing binary compounds have substantially smaller polymorph energy differences than their analogous Ca or Sr compounds. Additionally, MgTe has a WZ ground-state crystal structure while all other binary compounds investigated here have RS ground-state structures. To understand the origins of this difference, we utilized COHP analysis to quantify the bonding interactions within each of the binary structures. The COHP and DOS for each binary structure are shown in Figures S3-S6 of the SI. The net covalent bond energy ($\Sigma$), calculated from integration of the COHP and the Madelung energy for each structure are summarized in Figure 6. This analysis reveals that ionic/electrostatic interactions provide the major contribution to the bonding in the Ca and Sr chalcogenide materials, as indicated by their near-zero net covalency. Although Mg compounds are also largely dominated by ionic interactions, most Mg structures also have a small covalent contribution. As depicted in Figure 6a, the Madelung energy increases with decreasing anion radii and with decreasing cation radii, indicating more favorable electrostatic interactions for the more dense structures. The RS polymorph exhibits the largest Madelung energy for all compounds except for MgTe and SrO, and therefore indicates that the RS structures have the strongest ionic bonding of the polymorphs studied. This result matches our prediction that the RS polymorph is the most stable polymorph of all of the compositions considered except MgTe. However, the purely ionic contribution to the bonding fails to fully explain the different polymorph stability ordering of MgTe and SrO, where WZ and RS are the ground-state structures, but NA and BN have the largest Madelung energies, respectively.

Covalent bonding provides a weaker contribution to the overall bonding in these materials, but it is not negligible. This is most pronounced in the Mg chalcogenides, which have a broader DOS and minimal anti-bonding contribution near the Fermi energy (Figures S3-S6), indicating greater orbital overlap in these materials relative to the Ca and Sr compounds. For the non-RS Mg compounds, a positive net covalent bonding is observed (Figure 6b), stabilizing these polymorphs relative to the RS polymorph, which has zero net covalency. As the ionic bonding contribution decreases with increasing anion radius, the net covalent bonding becomes more important. As a result, the polymorph energy differences of the Mg compounds decrease with increasing anion radius. Ultimately, the WZ structure of MgTe, which has the largest net covalency of the MgTe structures, is stabilized over the RS polymorph, which has zero net covalency. Minimal covalent bonding is observed for all of the Ca chalcogenide polymorphs, and as a result, the RS polymorph is the most stable for all of the Ca compositions because it exhibits the largest Madelung energy. All of the non-RS polymorphs of the Sr chalcogenides exhibit net covalent antibonding ($\Sigma < 0$), destabilizing these compounds. The overall magnitude of this antibonding contribution decreases with increasing anion radii. As a result, the RS polymorphs for SrS, SrSe, and SrTe remain the most stable when considering both ionic and covalent bonding. However, this covalent destabilization does result in the RS polymorph becoming more stable than the BN polymorph for SrO, despite the BN polymorph exhibiting the larger Madelung energy.

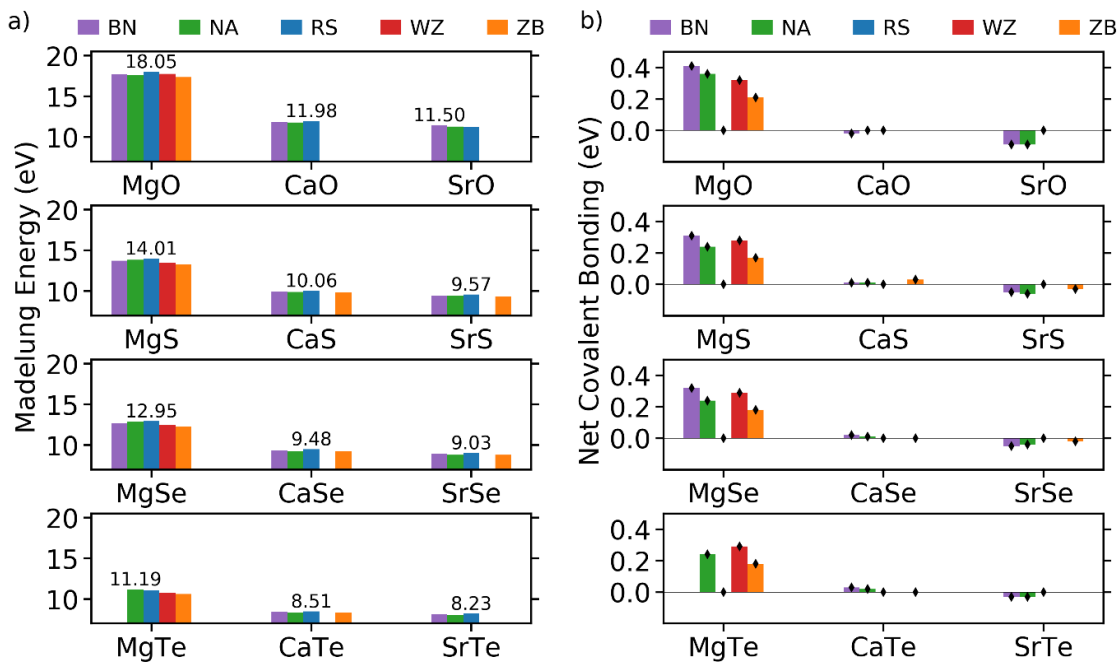

Figure 6. *Ionic and covalent bonding contribution in binary compounds.* a) Madelung energy of the cation calculated from the Bader charge for stable polymorphs. Anion charges are the same magnitude as the corresponding cation charge but are negative. Larger Madelung energies correspond to stronger ionic bonding. The value of the largest Madelung energy for each composition is given. The Madelung energy decreases with increasing cation and anion radii. b) Net covalent bond energy calculated from the integration of the COHP. Larger positive net covalent bonding corresponds to stronger covalent bonding. Negative net covalent bonding corresponds to destabilizing covalent interactions. The value of Σ is marked with a black diamond for the stable polymorphs. Non-RS Mg compounds have significantly larger Σ values than RS Mg polymorphs or any Ca or Sr polymorphs.

In summary, this work utilized GGA, meta-GGA, *GW*, and RPA calculations to comprehensively study the thermodynamic and electronic properties that result from anion alloying alkaline earth chalcogenides. We have shown that AEC alloys are largely dominated by the RS structure, but the small polymorph energy differences in the Mg-compounds enable a wide composition range for the ground-state WZ structure to be achieved through anion alloying between MgS and MgTe, which is not feasible through cation alloying within the AEC space. "Beyond-DFT" approaches proved necessary for modeling the Mg-chalcogenides and provide a qualitatively different picture of polymorph stability than DFT, changing not only the thermodynamic picture of the binary compounds, but also the alloy thermodynamics.

For all cations studied, anion alloying between S and Se results in low miscibility gap critical temperatures, thereby presenting a promising route for tuning these alloys' properties. Although alloying between S and Te requires non-equilibrium growth due the high miscibility gap temperatures, existing synthesis strategies, such as co-sputtering, offer a promising route for future experimental work. RS CaS$_{1-x}$Te$_x$ and SrS$_{1-x}$Te$_x$ were shown to possess wide band gaps and low hole effective masses, and WZ MgS$_{1-x}$Te$_x$ was shown to have a wide band gap and low electron effective mass, making them promising transparent materials. Based on the periodic trends in the alloy properties, it may be possible to synthesize (Se,Te) alloys at lower temperatures than (S,Te) alloys while maintaining their promising electronic properties. Our results provide guidance for future experimental studies to both target the improvement of WBG semiconductor materials properties for specific applications and synthesize new anion alloys based on the fundamental understanding our study provides for the origins of polymorph stability.

## 4 Methods

DFT computations of the alloy structures were performed within the Vienna Ab initio Simulation Package (VASP) using plane-wave periodic boundary conditions, the SCAN meta-GGA functional, a 650 eV planewave energy cutoff, a Γ-centered 4×4×4 $k$-point mesh, and projector augmented wave (PAW) pseudopotentials.[83-85] For consistency in the analysis between alloys, six valence electrons were explicitly described for chalcogenide elements while ten electrons were explicitly described for alkaline earth elements. Geometry relaxations were conducted using the conjugate gradient algorithm and converged to within $10^{-6}$ eV with each electronic self-consistency loop converged to within $10^{-7}$ eV.

The mixing enthalpies of the alloys were calculated using 48-atom supercell models of the crystal with random anion configurations generated using the special quasirandom structures (SQS) method for each composition and polymorph.[86] Nine phase mixtures approximately uniformly distributed across the binary end-member composition space were selected for the SQS compositions. The enthalpy of mixing, $\Delta H_m(x)$, at each composition was calculated according to Equation (1):

$$\Delta H_m(x) = E_{alloy} - xE_1 - (1-x)E_0 \quad (1)$$

where $E_{alloy}$ is the total energy of the alloy at alloy concentration $x$, and $E_0$ and $E_1$ are the total energies of the binary end-member compounds at $x = 0$ and $x = 1$, respectively. The alloy interaction parameters $\Omega$ and $\Omega'$ for each polymorph were determined by fitting $\Delta H_m(x)$ using a third-order polynomial given by Equation (2):

$$\Delta H_m(x) = xE_1 + (1-x)E_0 + \Omega x(1-x) + \Omega'(x^2 - x)(x - 0.5) \quad (2)$$

In this formalism, $\Omega$ causes the enthalpy of mixing curve to bow while $\Omega'$ shifts the enthalpy of mixing curve away from a symmetric parabola centered at $x = 0.5$. The free energy of mixing,

$\Delta G_m(T)$, which was used to construct $T$-$x$ phase diagrams, was calculated according to Equation (3):

$$\Delta G_m(x) = \Delta H_m(x) - T\Delta S_m(x) \qquad (3)$$

Where the entropy of mixing, $\Delta S_m(x)$, is the configurational entropy of the alloy as approximated by the regular solution model as given by Equation (4), where $k_b$ is Boltzmann's constant:

$$\Delta S_m(x) = -k_b[x \ln x + (1-x)\ln(1-x)] \qquad (4)$$

The miscibility gap (binodal line) was obtained from the common tangent construction of $\Delta G_m(x,T)$, and the spinodal gap was found from the locus of compositions that satisfy the condition $\delta^2 \Delta G_m(x)/\delta x^2 = 0$.

We note that for the alloy phase diagrams in this work shown in of Figures 1, 4, and S1, the state-of-matter phase transitions from the solid state that will occur for the sulfur-oxygen and sulfur-tellurium alloys within the plotted temperature range are not included in this evaluation. For instance, the liquid phases are not shown in the phase diagrams shown in Figures 1b and f, although $CaS_{1-x}O_x$ and $CaS_{1-x}Te_x$ are expected to melt within the temperature range considered.[87] Still, the simple regular solution model we employ serves as a measure of the scale of metastability. The region shown in grey beneath the spinodal curve is the region of spinodal decomposition where the free energy of mixing has negative curvature ($\delta^2 \Delta G_m(x)/\delta x^2 < 0$). In this region, any microscopic fluctuation in composition results in the spontaneous decomposition of the single-phase alloys to their constituent phases given sufficiently low kinetic barriers for diffusion. In the region shown in white between the spinodal and binodal curves, the alloy is metastable and the system tolerates small fluctuations in composition. In contrast to the spinodal decomposition region, the free energy of the system in the metastable region only decreases by a process of nucleation and growth accompanied by a large fluctuation in composition. In the regions shown in blue above the binodal curves, the single-phase alloys are fully thermodynamically miscible.

Due to the small energy differences between polymorphs of Mg-containing compounds, polymorphs of MgO, MgS, MgSe, and MgTe were computed using RPA within the adiabatic-connection fluctuation-dissipation theorem (RPA-ACFDT).[63] All RPA calculations were performed in VASP using the Perdew-Burke-Ernzerhof (PBE) generalized gradient approximation exchange-correlation functional to compute the input wavefunction and geometry.[57,83,88] RPA calculations utilized PAW pseudopotentials with a 520 eV plane wave cutoff and a 350 eV response function cutoff.[84] Several RPA calculations below this response function cutoff were performed to extrapolate the RPA correlation energy to infinite cutoff.[63] A constant number of bands were used for each material: 216 bands/atom for MgO, 432 bands/atom for MgS, 540 bands/atom for MgSe, and 648 bands/atom for MgTe. All calculations were spin un-polarized. All structures were calculated with a Γ-centered *k*-point mesh density of at least 1000 *k*-points per reciprocal atom. The minimum energy of each polymorph and the corresponding volume, bulk modulus, and lattice parameters were determined by fitting the third-order Birch-Murnaghan equation of state to the RPA total energies of several strained lattices for each polymorph.[89] Each strained lattice was first relaxed with PBE while holding the volume fixed to allow for anisotropic effects. Polymorph energy differences were further refined for the Mg compounds by including the zero-point energy (ZPE) of each polymorph. The ZPE was computed using density functional perturbation theory with the structure determined to have the minimum energy using RPA for each polymorph.[90] The reported 0 K alloy enthalpies of mixing for Mg compounds were calculated by adjusting the SCAN-predicted alloy mixing enthalpy at *x* by a linear weighting of the difference between the end-member SCAN and RPA-calculated energies: $\Delta H_{m,RPA,x} = \Delta H_{m,SCAN,x} + (xA + (1-x)B)$, where the weights are given by $A = \Delta H_{m,RPA,x=1} - \Delta H_{m,SCAN,x=1}$ and $B = \Delta H_{m,RPA,x=0} - \Delta H_{m,SCAN,x=0}$.[91]

The SCAN meta-GGA functional has been shown to predict more accurate band gaps than PBE, but SCAN band gaps are still expected to be systematically underestimated.[62,92,93] While the *GW* approximation has been shown to significantly improve the predicted band gaps of materials, particularly those with a heavy element (Z > 29), it is computationally prohibitive to perform these calculations for the many large supercells required to model alloys.[94,95] Although the band gaps are underestimated by DFT, the changes in the electronic structure across the composition range are well-captured by DFT in these materials.[96] Therefore, we computed the theoretical band gaps of the alloys through a linear scaling of the DFT-predicted band gaps to the end-member band gaps calculated within the *GW* approximation[97] where we calculated the "effective *GW*" band gaps using the simple linear scaling, as described previously.[91] A similar *GW*-corrected linear weighting was also utilized for the reported density of states (DOS) effective masses. The results of the *GW* approximation are publicly available through the National Renewable Energy Laboratory (NREL) materials database for all materials except CaSe and CaTe, which we calculated using input parameters consistent with those used to compute those tabulated in the NREL database.[98-100] For each compound, the optical bowing coefficient (*b*), which provides a measure of the deviation in the band gap of an alloy from Végard's law, is determined by fitting the computed band gaps across the composition range, as given by Equation (5):[101,102]

$$E_g(x) = xE_{g,x=1} + (x-1)E_{g,x=0} - bx(1-x) \tag{5}$$

The scatter in the calculated effective masses and band gaps shown in Figures 2 and 5 largely results from a finite size effect due to the utilization of a 48-atom SQS periodic representation of the random alloys used to keep these calculations tractable; however, the computed trends shown here are expected to be representative of the change in the band gap across the composition space. This effect has been demonstrated and discussed in detail in the literature.[103] The noise in these results can be reduced by using larger supercell models of the alloys to allow for a more thorough sampling of the possible local alloy configurations. Noise reduction will be important for future studies that aim to fully optimize a property in each compositional space towards a target application.

Figure 7 shows the five different crystal structures considered for this study: WZ, BN, ZB, RS, and NA. These structures differ in their stacking (face-centered cubic in RS and ZB, hexagonal close-packed in WZ, BN, and NA) and their cation coordination (tetrahedral in ZB and WZ, trigonal bipyramidal in BN, and octahedral in RS and NA). ZB structures of CaO, SrO, and SrTe, WZ structures containing either Ca or Sr, and BN structures of MgTe were calculated to be over 400 meV/formula unit (fu) higher in energy than the ground-state structure and to have no kinetic barrier to transition into the RS, BN, and WS structures, respectively. Consequently, alloys involving those structures were not computed. End-member energies of those structures were calculated using a fixed-volume relaxation to ensure they retained the desired crystal structure.

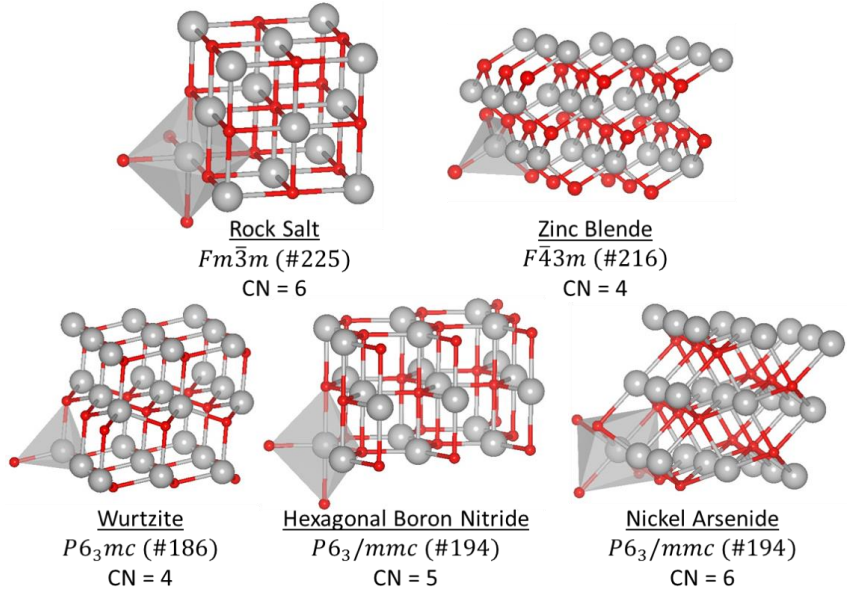

Figure 7. *Computed polymorphs.* The five polymorphs of the alkaline-earth chalcogenide alloys computed in this work illustrated as ball-and-stick arrays of anions (red) and cations (grey). The space group and coordination number (CN) are given for each structure. Rock salt and zinc blende are face-centered cubic structures while wurtzite, hexagonal boron nitride and nickel arsenide are hexagonal structures. Wurtzite and zinc blende have tetrahedral cation coordination, hexagonal boron nitride has trigonal bipyramidal cation coordination, and rock salt and nickel arsenide have octahedral coordination.

Crystal Orbital Hamilton Populations (COHP),[104] as implemented in the LOBSTER code,[105] were used to quantify the bonding interactions within the materials studied. To normalize the comparison of COHPs between polymorphs and across compositions, the calculated energy levels for each polymorph were aligned with respect to their core levels and the populations were normalized per valence electron. The net covalent bond energy, $\Sigma$, was obtained by the energy-weighted integral of COHP states below and near the Fermi energy:[106,107]

$$\Sigma = \int_{-15eV}^{\varepsilon_F} -COHP(E)E\,dE \qquad (6)$$

where $\varepsilon_F$ is the Fermi energy and $E$ is the core-level-aligned energy. Madelung energies were computed from Bader charge densities[108,109] for end-member polymorphs using the Ewald summation approach[110] as implemented in pymatgen[111] to quantify the electrostatic interactions in these materials.

## 5  Data Availability

The data to support the findings of this study are available from the corresponding author upon request.

# 7    Acknowledgements

CBM and SLM acknowledge support from the U.S. Department of Energy's Office of Energy Efficiency and Renewable Energy, Fuel Cells Technologies Office Award No. DE-EE0008088, "Computationally Accelerated Discovery and Experimental Demonstration of High-Performance Materials for Advanced Solar Thermochemical Hydrogen Production." CBM, CJB, and JMC also acknowledge support from NSF CHEM 1800592, CBET 1806079 and CBET 2016225. AMH acknowledges funding from the Office of Science (SC), Office of Basic Energy Sciences (BES), as part of the Energy Frontier Research Center "Center for Next Generation of Materials Design: Incorporating Metastability" under contract DE-AC36-08GO28308. NRS was supported by a U.S. Department of Education Graduate Assistance in Areas of National Need Fellowship under the Materials and Energy Conversion and Sustainability Program. This research was performed using computational resources sponsored by the Department of Energy's Office of Energy Efficiency and Renewable Energy and located at the National Renewable Energy Laboratory. The views expressed in the article do not necessarily represent the views of the DOE or the U.S. Government.

# 8  Author Contributions

S.L.M performed the thermodynamic calculations and drafted the manuscript. J.M.C. performed the RPA calculations and assisted in writing the manuscript. C.J.B and N.R.S. performed the COOP analysis and assisted in writing the manuscript. A.M.H. and C.B.M supervised the project, provided guidance on the calculations and analysis and assisted in writing the manuscript. All authors commented on the results and reviewed the manuscript.

**Supplemental Information**

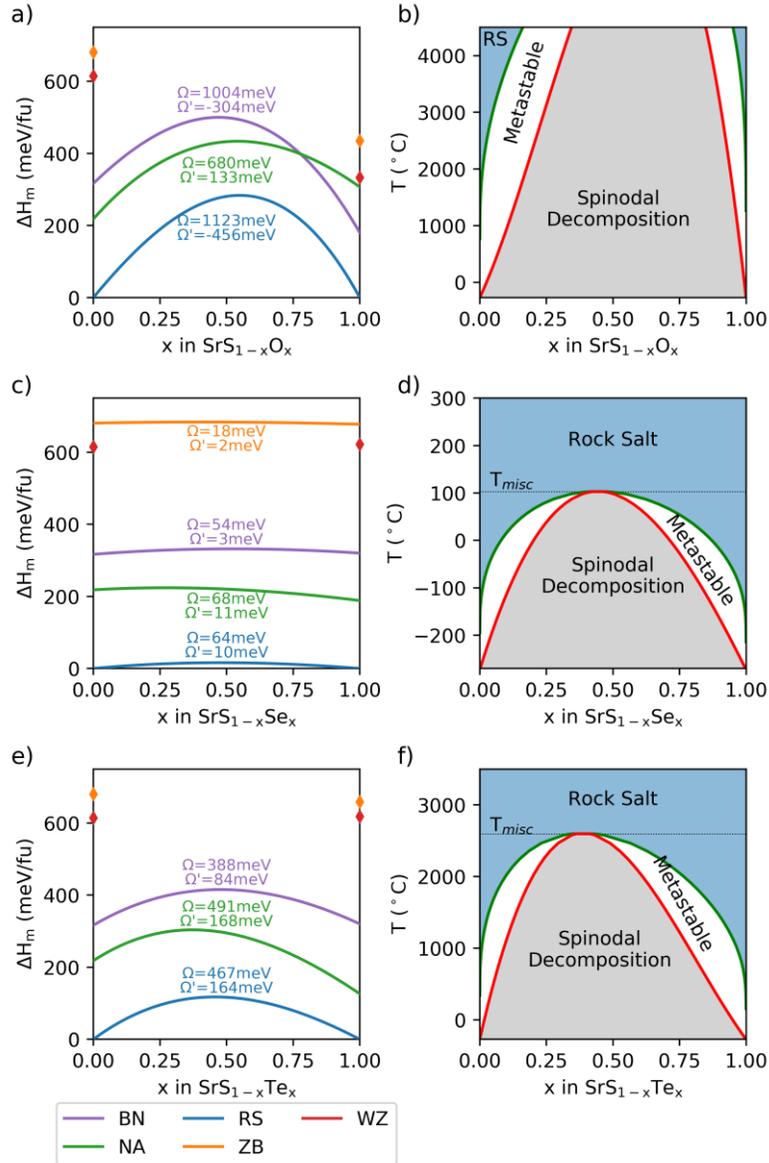

Figure S1. *Thermodynamic stability of Sr compounds.* Calculated mixing enthalpy curves for each of the five polymorphs considered (left) and resulting temperature-composition phase diagram (right) for strontium-containing alloys a-b) $SrS_{1-x}O_x$ c-d) $SrS_{1-x}Se_x$, and e-f) $SrS_{1-x}Te_x$. In the mixing enthalpy figures alloy interaction parameters $\Omega$ and $\Omega$' are given for each polymorph. For compounds found to be unstable, only the end-member energies are displayed. In the phase diagrams the spinodal line (red) and binodal line (green) separate the regions of stability (blue), metastability (white), and instability (grey).

Table S1. RPA calculated structural parameters of MgO, MgS, MgSe, and MgTe for the hexagonal boron nitride (BN), nickel arsenide (NA), rock salt (RS), wurtzite (WZ), and zinc blende (ZB) structures. Structural parameters include the *a* lattice parameter and bulk modulus (B and B') for all structures, and the *c* lattice parameter and internal *u* parameter for hexagonal structures.

|      |    | $a$ (Å) | $c$ (Å) | $u$  | B (Mbar) | B'    |
|------|----|---------|---------|------|----------|-------|
| MgO  | BN | 3.474   | 4.186   | 0.50 | 0.62     | 4.39  |
|      | NA | 2.926   | 5.108   | 0.25 | 0.68     | 3.50  |
|      | RS | 4.215   | --      | --   | 0.69     | 3.32  |
|      | WZ | 3.477   | 4.189   | 0.50 | 0.61     | 1.91  |
|      | ZB | 4.540   | --      | --   | 0.63     | 9.54  |
| MgS  | BN | 4.252   | 5.219   | 0.50 | 0.22     | -3.02 |
|      | NA | 3.610   | 6.096   | 0.25 | 0.38     | 7.36  |
|      | RS | 5.152   | --      | --   | 0.35     | 3.45  |
|      | WZ | 3.990   | 6.339   | 0.38 | 0.28     | 5.41  |
|      | ZB | 5.598   | --      | --   | 0.26     | 3.68  |
| MgSe | BN | 4.472   | 5.537   | 0.50 | 0.24     | 4.17  |
|      | NA | 3.821   | 6.371   | 0.25 | 0.31     | 6.16  |
|      | RS | 5.433   | --      | --   | 0.29     | 4.88  |
|      | WZ | 4.186   | 6.727   | 0.38 | 0.24     | 9.99  |
|      | ZB | 5.894   | --      | --   | 0.23     | 4.05  |
| MgTe | BN | 4.835   | 6.103   | 0.50 | 0.17     | 3.79  |
|      | NA | 4.171   | 6.789   | 0.25 | 0.22     | 4.63  |
|      | RS | 5.898   | --      | --   | 0.22     | 3.84  |
|      | WZ | 4.531   | 7.372   | 0.38 | 0.17     | 3.85  |
|      | ZB | 6.400   | --      | --   | 0.16     | 3.83  |

**Supplemental Discussion of Periodic Trends**

Examining the results for the nine alloy systems studied in this work reveals several trends. First, $\Omega$ decreases across the cationic series Mg > Ca > Sr, as shown in Figure S2a. Because the ionic radii of Mg, Ca, and Sr in an octahedral coordination environment are 0.72, 1.0, and 1.18 Å, respectively, the volumes of the resulting binary compound will also increase when these cations are bound to the same anion species. For example, the calculated volumes of MgS, CaS, and SrS in the RS structure are 35.1, 46.4, and 55.1 Å$^3$/fu, respectively. As a result, compounds with the smallest cell volumes will be least amenable to alloying with larger anions. Indeed, this is consistent with our results that show the (S,O) alloys exhibit greater $\Omega$ values than the (S,Se) or (S,Te) alloys. Second, we found that $\Omega$ decreases across the anionic series (S,O) > (S,Te) > (S,Se) (Figure S2a). This trend can be explained by comparing the difference in radii of each anion pair following an analogous argument as the cation trend. In an octahedral coordination environment, the differences in the ionic radii for these anion pairs are 0.44, 0.37, and 0.14 Å for (S,O), (S,Te), and (S,Se), respectively. As the difference between the ionic radii increases, structural distortions increase upon alloying. These distortions then disrupt the electronic structure, increasing the energetic penalty for alloying, and thus strengthen the driving force for phase separation. Although not shown in Figure S2, our results for $\Omega'$ show that the mixing enthalpy curve is shifted towards the more electronegative anion (O > S ≈ Se > Te) and the magnitude of the shift depends on the magnitude of the difference in electronegativity. The direction of the shift of the enthalpy of mixing curves are also reflected in the stability (as defined by the calculated formation enthalpy) of the binary AECs, again following the ordering AEO > AES ≈ AESe > AETe.

Both the cation and anion trends exhibited in $\Omega$ are also reflected in the miscibility temperature trends for these alloys

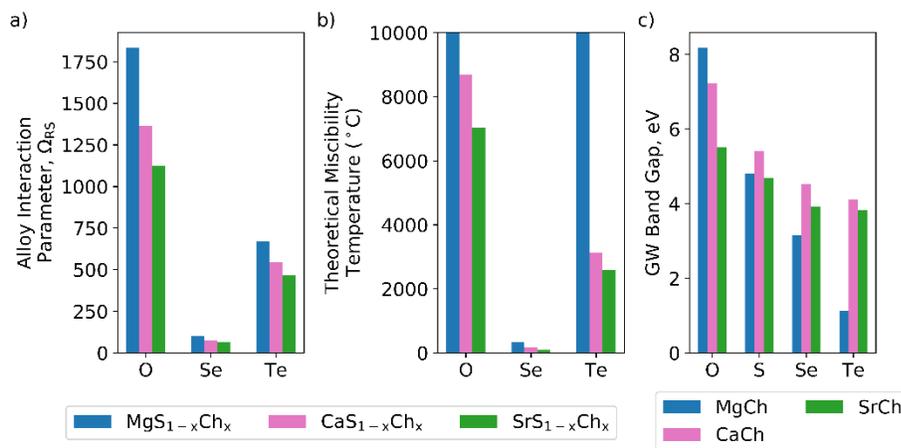

Figure S2. *Periodic trends across alloys.* Periodic trends in the a) alloy interaction parameter of the alloys, b) theoretical miscibility temperature of the alloys, and c) *GW* computed band gaps of the binary end-member compounds.

(Figure S2b), highlighting how understanding these trends will aid in future synthesis of these compounds. $MgS_{1-x}Te_x$ is a notable exception to this however, because we predict that it is a heterostructural alloy between the RS and WZ crystal structures and therefore exhibits an infinite theoretical miscibility gap critical temperature at the phase transition point. The lowest alloy interaction parameters across all alloys considered are observed for the $SrS_{1-x}Se_x$ system for which $\Omega$ = 64 meV for the ground-state RS structure, which produces a miscibility gap critical temperature of just 105 °C. The highest alloy interaction parameters we predicted for an isostructural alloy are for the $CaS_{1-x}O_x$ system ($\Omega_{RS}$ = 1365 meV), resulting in a remarkably narrow miscibility region and a theoretical miscibility temperature of over 8700 °C.

The computed band gaps of the Mg, Ca, and Sr binary chalcogenides decrease with increasing anion radii from O to Te, as shown in Figure S2c. This trend is expected given the electronegativity of these elements, which also decreases moving down the periodic table from O to Te. As the electronegativity increases, the bond strength increases and the energy difference between the bonding and antibonding states grows, increasing the band gap. This results in calculated band gaps as high as 5.7 eV for RS MgO and as low as 1.0 eV for RS MgTe.

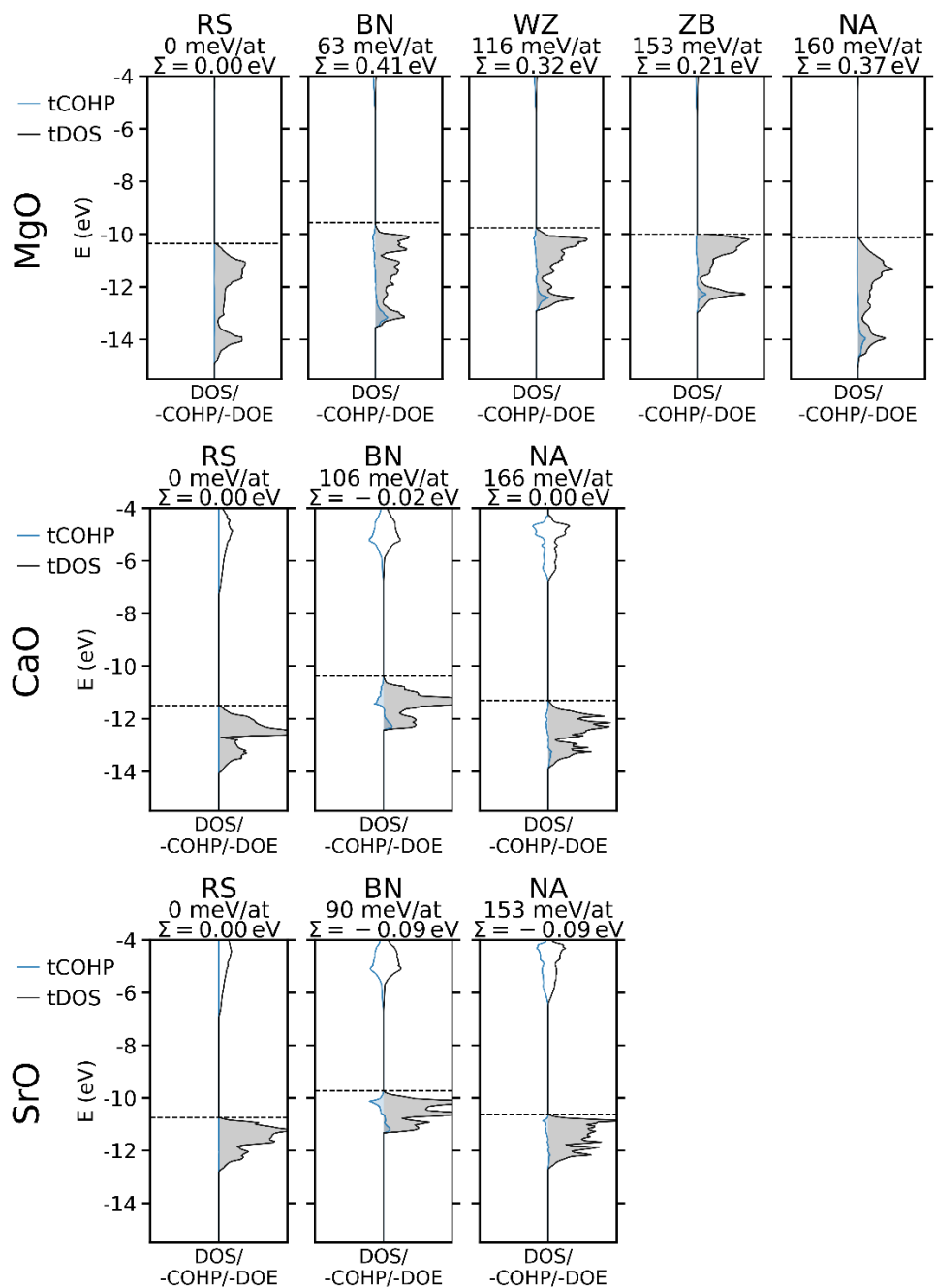

Figure S3. -COHP (blue) and DOS (grey) for each stable oxide structure studied ordered from left to right by stability. The COHP is plotted such that areas to the right of 0 indicate bonding and those to the left indicate antibonding. The SCAN polymorph energy and net covalent bond energy ($\Sigma$) is given for each structure. The Fermi energy is shown as a dashed line.

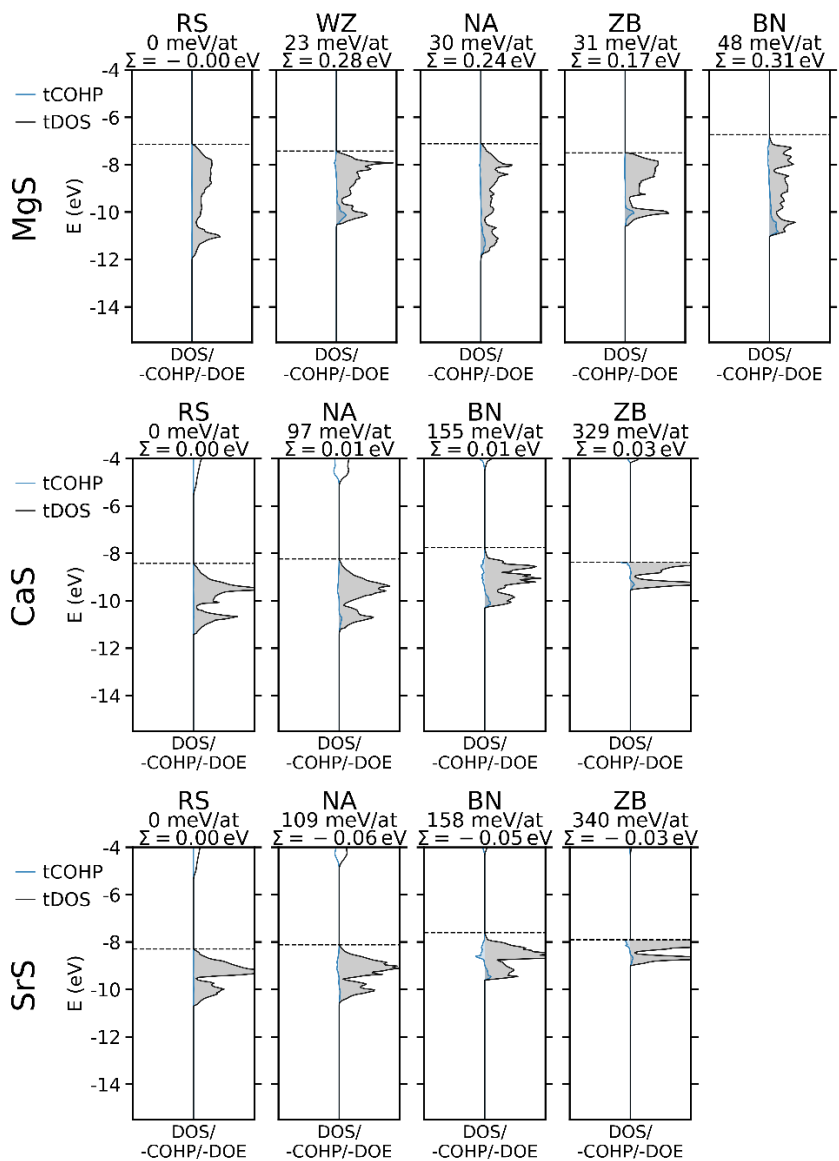

Figure S4. -COHP (blue) and DOS (grey) for each stable sulfide structure studied ordered from left to right by stability. The COHP is plotted such that areas to the right of 0 indicate bonding and those to the left indicate antibonding. The SCAN polymorph energy and net covalent bond energy ($\Sigma$) is given for each structure. The Fermi energy is shown as a dashed line.

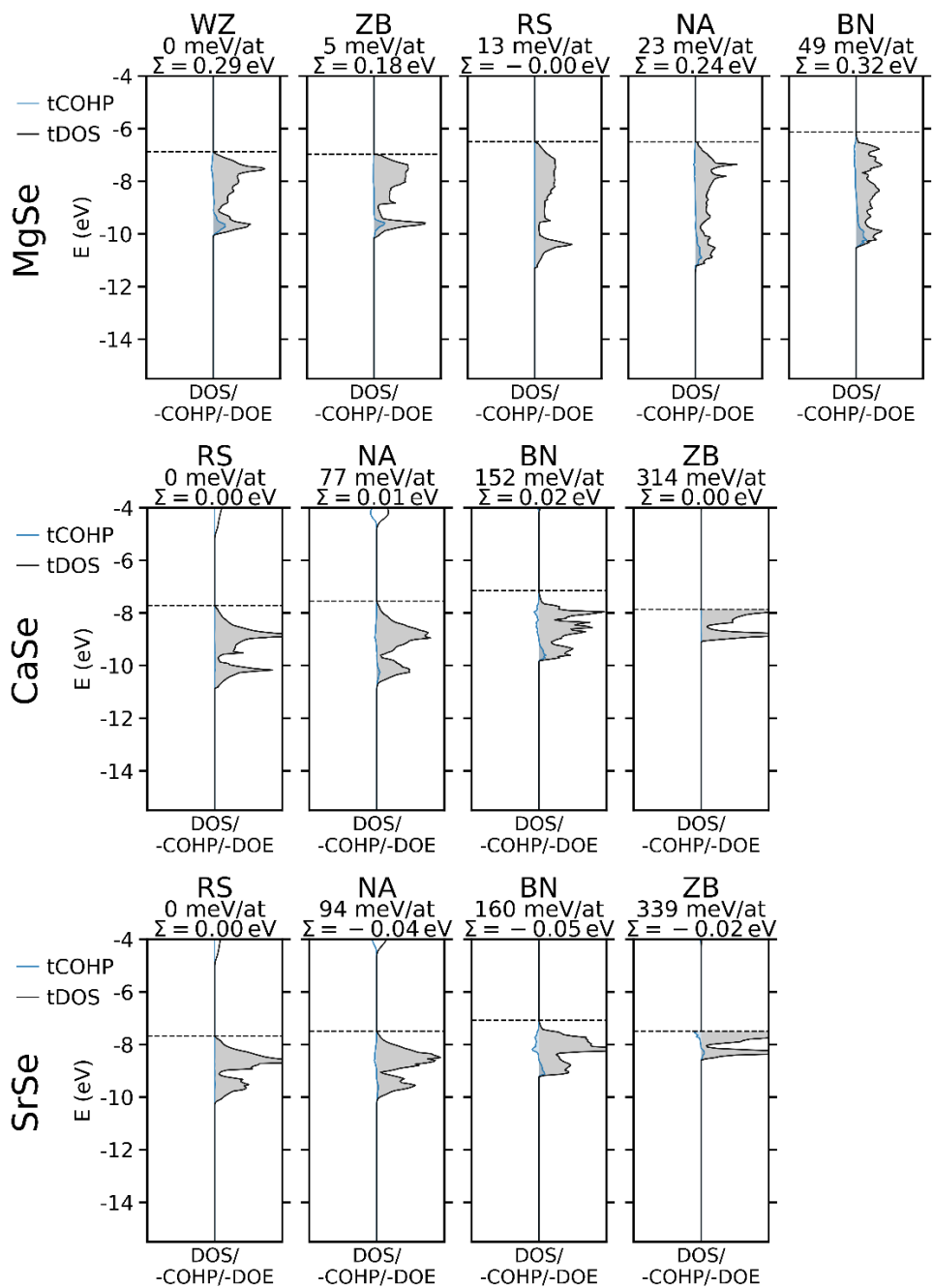

Figure S5. -COHP (blue) and DOS (grey) for each stable sellenide structure studied ordered from left to right by stability. The COHP is plotted such that areas to the right of 0 indicate bonding and those to the left indicate antibonding. The SCAN polymorph energy and net covalent bond energy ($\Sigma$) is given for each structure. The Fermi energy is shown as a dashed line.

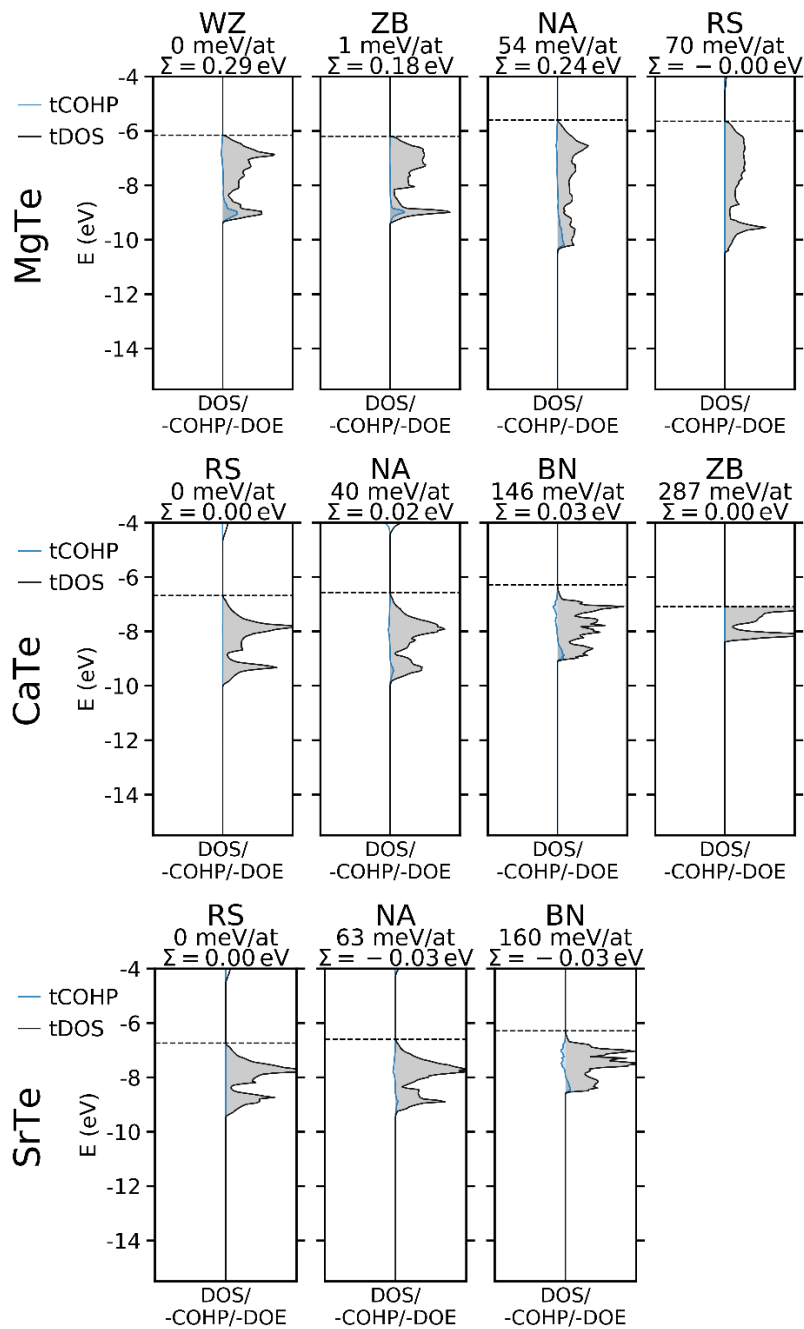

Figure S6. -COHP (blue) and DOS (grey) for each stable telluride structure studied ordered from left to right by stability. The COHP is plotted such that areas to the right of 0 indicate bonding and those to the left indicate antibonding. The SCAN polymorph energy and net covalent bond energy ($\Sigma$) is given for each structure. The Fermi energy is shown as a dashed line.